\documentclass[twocolumn]{paper}
\usepackage{indentfirst}
\usepackage{amsmath}
\usepackage{graphicx}
\usepackage{nomencl}
\usepackage{SIunits}
\usepackage{color}
\usepackage{booktabs}

\newcommand{\rc}{\reciprocal}

\title{Guidelines for the Calculation of Bound Molecular Spectra}

\author{M. Lino da Silva}

\institution{Centro de F\'{i}sica de Plasmas, Instituto
Superior T\'{e}cnico, Av. Rovisco Pais, 1049-001 Lisboa, Portugal\\%
\texttt{mlinodasilva@mail.ist.utl.pt}/Fax: +351 21 841 90 13}

\begin{document}

\maketitle

\begin{abstract}
Line-by-line calculations are becoming the standard procedure for
carrying spectral simulations. However, it is important to insure
the accuracy of such spectral simulations through the choice of
adapted models for the simulation of key parameters such as line
position, intensity, and shape. Moreover, it is necessary to rely
on accurate spectral data to guaranty the accuracy of the
simulated spectra. A discussion on the most accurate models
available for such calculations is presented for diatomic and
linear polyatomic discrete radiation, and possible reductions on
the number of calculated lines are discussed in order to reduce
memory and computational overheads. Examples of different
approaches for the simulation of experimentally determined
low-pressure molecular spectra are presented. The accuracy of
different simulation approaches is discussed and it is verified
that a careful choice of applied computational models and
spectroscopic datasets yields precise approximations of the
measured spectra.
\end{abstract}



\section{Introduction}

Advances in the capabilities of modern day computing have allowed
carrying line-by-line calculations in a systematic manner for the
field of spectroscopy modelling. Various applications such as
species concentrations and temperature measurements in low
pressure plasmas \cite{Laux:2001,Laux:2003}, or the calculation of
radiative fluxes for atmospheric entry flows \cite{Park:2004},
greatly benefit from such techniques. However, the accuracy of
such spectral simulations may vary consequently, depending on the
methods used for line-by-line calculations, but also on the
applied spectroscopic datasets.

The different methods used for the calculation of key spectral
parameters such as line positions, intensities, and shapes are
discussed in this paper. Available state of the art models for
each of these parameters are presented, and some simplifications
to such spectral models are discussed, leading to lower memory
requirements and calculation times for the used computing systems.
The models presented and discussed in this paper are valid for
diatomic rovibronic transitions, but also for linear polyatomic
rovibrational transitions (such as transitions from the CO$_{2}$
molecule). The second part of this work presents some simulations
of experimentally measured high resolution spectra issued from low
pressure and high enthalpy plasmas. The discrepancies that may
derive from the selection of different simulation models and
spectral datasets will be highlighted through the comparison of
different simulated spectra with the measured spectra. It will be
verified that a careful selection of adequate models, linked to a
selection of accurate spectroscopic data, may yield a very
accurate reproduction of high resolution measured spectra.

\section{Theoretical Models for the Simulation of Bound Spectra
in the Line-by-Line Approach}

Discrete molecular radiation can be characterized unambiguously
through three parameters: line position, intensity, and shape

Determination of line positions depends on the quantification of
the energy levels of the molecule bound states. Line intensities
depend on the probabilities of transition between the different
states, as well as on the population of these states. Finally,
line shapes depend on the local conditions of the gas in which the
transition takes place\footnote{line positions shifts can also
occur from broadening mechanisms}.

An overview of the different methods and approaches to several
levels of accuracy will be presented in this section for these
three parameters.

\subsection{Line Positions Calculations}

Three different approaches exist for the calculation of quantum
level energies which will determine line positions of bound
molecular spectra:

The first one allows a broad calculation of any number of
vibrational and rotational levels for a given electronic
transition of a molecule. Calculations are performed using
equilibrium constants, which give the vibrational and rotational
constants for the given electronic transition in a broad range of
vibration and rotation levels.

The second one uses level-by-level spectroscopic constants and
allows the calculation of any number of rotational levels for
given electronic and vibrational transition levels of a molecule,
as the band origin and rotational constants are set for each
vibration level. This approach generally allows a better
determination of specific vibration levels energies but prevents
one from simulating further levels than those for which
spectroscopic data is available.

The third one, and the more accurate, requires diagonalising the
corresponding hamiltonian matrix for each rovibronic state. This
approach can be useful when the experimental spectra to be
simulated is strongly perturbed although it leads to larger
computational times.

\subsubsection{Line position calculations using equilibrium constants in matrix form}

Klein-Dunham coefficients allow a clear and unambiguous
determination of level positions, compared to the traditional
spectroscopic developments, prone to confusions and errors (for
instance, the parameter $\gamma_{e}$, useful for the calculation
of the rotational constant $B_{v}$ can sometimes be confused with
the spin-rotation interaction coefficient $\gamma$).

Level positions (in \centi\rc\metre) are calculated according to
the following relation:

\begin{equation}
E_{e,v,J}=\sum_{i,j}Y_{ij}\left(v+1/2\right)^{i}[F(J)]^{j},
\label{eq:EevJ}
\end{equation}

The formalism proposed here is consistent with the effective
Hamiltonian of Zare \cite{Zare:1973}. Other formalisms may exist,
and care should be exercised when using published
coefficients.\bigskip

Traditional spectroscopic developments are related to Zare's
Klein-Dunham coefficients by:

\begin{subequations}
\begin{align}
\label{eq:KleinDunham1}
G_{v}&=\sum_{i=0,\,\ldots}Y_{i0}\left(v+\frac{1}{2}\right)^{i}=%
T_{e}+\omega_{e}\left(v+\frac{1}{2}\right)\\
&-\omega_{e}x_{e}\left(v+\frac{1}{2}\right)^{2}
+\omega_{e}y_{e}\left(v+\frac{1}{2}\right)^{3}+\ldots\notag\\
\label{eq:KleinDunham2}
B_{v}&=\sum_{i=0,\,\ldots}Y_{i1}\left(v+\frac{1}{2}\right)^{i}=%
B_{e}-\alpha_{e}\left(v+\frac{1}{2}\right)\\
&+\gamma_{e}\left(v+\frac{1}{2}\right)^{2}+\ldots\notag\\
\label{eq:KleinDunham3}
D_{v}&=-\sum_{i=0,\,\ldots}Y_{i2}\left(v+\frac{1}{2}\right)^{i}=%
D_{e}+\beta_{e}\left(v+\frac{1}{2}\right)+\ldots\\
\label{eq:KleinDunham4}
H_{v}&=-\sum_{i=0,\,\ldots}Y_{i3}\left(v+\frac{1}{2}\right)^{i},\ldots%
\end{align}
\end{subequations}

The use of Klein-Dunham expansions, unlike traditional
developments, may also prevent situations were neglecting higher
order corrections leads to considerable shifts for the position of
calculated lines compared to the experimental spectrum
\cite{Laux:1993}.

Setting a matrix of Klein-Dunham coefficients such as $i=10$ and
$j=7$ in computer routines for line position calculations suffices
in order to account for the available polynomial expansions as
verified by a broad review of available spectroscopic coefficients
by the author.

For multiplet transitions, the expression for the level energies
differs slightly from Eq. \ref{eq:EevJ}, but the general form of
the Klein--Dunham matrix can be used. Expressions for the
different multiplet level energies are presented in appendix
\ref{sec:LevEn}.

\subsubsection{Further line position calculation methods}

When the level positions can no longer be accurately approximated
through Klein-Dunham expansions (as when vibrational perturbations
of the spectra are present), spectroscopic constants for each
vibrational level must be used. Level spectroscopic constants
obtained from fits of the rotational lines for each vibrational
band are given in this case. Using such level constants usually
results in more accurate predictions of the level energies.
However calculations are restricted to the vibrational levels were
spectroscopic constants are available, unlike Klein-Dunham
expansions which allow higher level extrapolations of the
available spectroscopic data.\bigskip

When perturbations are present in the spectra, the polynomial
expansions described previously no longer suffice for the accurate
simulation of line positions. Instead, one has to solve the
Hamiltonian matrix, taking into account the effects of the
perturbing states using the perturbation method
\cite{Lefebvre-Brion:1986}. This leads to very precise
calculations of the line positions (typically less than 0.1
\centi\rc\metre). However, this method requires the calculation of
the proper values of a $n*n$ matrix for each rovibronic state
where $n$ is the level multiplicity.

\subsection{Line Intensities Calculations}
\label{sec:rovibronic}

An overview of the different methods used for the calculation of
line emission and absorption as well as the different difficulties
and approximations seldom encountered will be presented in this
section.

The emission coefficient for a single line (excluding broadening
mechanisms) is calculated according to the relation

\begin{align}
\varepsilon_{\nu}&=\frac{N_{u}A_{ul}\Delta E}{4\pi}\notag\\%
                 &=\frac{1}{4\pi}N_{e'v'J'}A^{e'\phantom{'}v'}_{e''v''}\frac{S^{\Lambda'\phantom{'}J'}_{\Lambda''J''}}{2J+1}h\nu%
\label{eq:emicoeffdi}%
\end{align}

Eq. \ref{eq:emicoeffdi} highlights the additional difficulties
related to line intensity calculations. These depend not only on
the line positions (through the accounting of the transition
energy $\Delta E$), but also on the transition probabilities
$A_{ul}$ and the number density of the initial state $N_{u}$.
Although the two former quantities only depend on the transition
parameters, being calculated according to quantum mechanics laws,
the latter depends on the state of the studied gas.

\subsubsection{Dependence between line emission and absorption coefficients}

The radiative properties of a gas can be unambiguously known
through the determination of it's wavelength-dependent emission
and absorption coefficients $\varepsilon_{\nu}$ and $\alpha(\nu)$.
These two quantities are not independent however, and line
absorption coefficients can be determined from the line emission
coefficients.\bigskip

The absorption coefficient including spontaneous and induced
absorption (adopting the normalization factor $2J+1$ for the
H\"{o}nl-London coefficients and excluding broadening mechanisms)
is given by the relation

\begin{equation}
\alpha(\nu)=\frac{h\nu}{c}\left(N_{l}B_{lu}-N_{u}B_{ul}\right)%
\label{eq:abscoeffdi}%
\end{equation}

In complete thermodynamic equilibrium (CTE), the emission and
absorption coefficients are related through Planck's law:

\begin{equation}
\frac{\varepsilon_{\nu}^{0}}{\alpha^{0}(\nu)}=L_{\nu}^{0}=2hc\nu^{3}\left[\exp\left(\frac{h\nu}{kT}\right)-1\right]^{-1}
\label{eq:Planck}%
\end{equation}

One can therefore deduce the following relationships between the
Einstein spontaneous emission coefficient $A_{ul}$, the Einstein
spontaneous absorption coefficient $B_{lu}$, and the Einstein
induced absorption coefficients $B_{ul}$:

\begin{align*}
&\frac{A_{ul}}{B_{ul}}=8\pi h\nu^{3} & \textrm{and} & &
g_{l}B_{lu}=g_{u}B_{ul}
\label{eq:A_B}%
\end{align*}

As these relationships hold even in non-local thermodynamic (NLTE)
conditions, after simple algebraic manipulation, one can relate
the line emission and absorption coefficients (in wavenumber
units) through the expression

\begin{equation}
\alpha(\bar{\nu})=\varepsilon_{\bar{\nu}}\frac{1}{2hc^{2}\bar{\nu}^{3}}\left(\frac{g_{l}n_{l}}{g_{u}n_{u}}-1\right)
\label{eq:relemiabscoeff}%
\end{equation}

which means that only one of these two coefficients needs to be
calculated.\bigskip

For the more restrictive case of a Boltzmann distribution of the
internal levels populations where
\hbox{$n_{i}=g_{i}\exp\left(-\frac{E_{i}}{kT}\right)/Q_{tot}$} one
recovers the relation between the emission and absorption
coefficient defined by Planck's Law (Eq. \ref{eq:Planck}).

\subsubsection{Determination of the initial quantum levels populations $N_{u}$}

In thermodynamic equilibrium, the population of a quantum level
can be calculated straightforwardly from the well-known Boltzmann
equilibrium relation

\begin{equation}
\frac{N_{i}}{N}=\frac{Q_{i}}{\sum_{i}Q_{i}}
\end{equation}

with
$Q_{i}=g_{i}\exp\left(-\frac{E_{i}}{k_{B}T_{i}}\right)$.\bigskip

However, accurate calculations of atomic and molecular species
partition functions requires a set of accurate spectroscopic
constants up to higher quantum levels (close to the dissociation
limits for molecules and ionization limits for atoms). Namely, the
lowering of the ionisation threshold for atomic species has to be
considered in the plasma state, when including the contribution of
the atomic Rydberg states to the overall partition function
\cite{Giordano:1994}. Also, for molecular partition functions
calculations, it is necessary to determine the maximum rovibronic
levels which can be achieved before the dissociation of the
molecule occurs (including superdissociative states).\bigskip

Out of thermodynamic equilibrium, no straightforward method for
calculating the levels populations exists, and one has to resort
to state-to-state models which explicitly take into account the
different possible discrete states of a gas species. The
development of accurate state-to-state models has been carried by
different research teams, and the reader should refer to the
references
\cite{Capitelli:2000,Chernyi:2002,Sarrete:1995,Guerra:2004,Corse:1982,Capitelli:1997,Armenise:1999}
for a more detailed description of such models.

\subsubsection{Determination of the transition probabilities
$A_{ul}$}

As it was seen before, the global transition probability can be
decomposed into a vibronic and rotational part:

\begin{equation}
A_{ul}=A^{e'v'}_{e''v''}\cdot A^{\Lambda'J'}_{\Lambda''J''}
\end{equation}

The rotational transition probability is calculated according to
the H\"{o}nl-London factors which depend on the electronic
transition type ($^{n}\Lambda\leftrightarrow\!\!^{n}\Lambda$).

\begin{equation}
A^{\Lambda'J'}_{\Lambda''J''}=\frac{S^{\Lambda'J'}_{\Lambda''J''}}{2J'+1}
\end{equation}

in which the normalisation rule reads

\begin{equation}
\sum_{J''}S^{\Lambda'J'}_{\Lambda''J''}(J')=(2J'+1)
\end{equation}

Some authors \cite{Whiting:1980} use a slightly normalisation rule

\begin{equation}
\sum_{J''}S^{\Lambda'J'}_{\Lambda''J''}(J')=(2-\delta_{0,\Lambda'})(2S+1)(2J'+1)
\end{equation}

Expressions of the H\"{o}nl-London factors for the different types
transitions are given by several authors and can be found in Refs.
\cite{Budo:1937,Schadee:1964,Arnold:1969,Kovacs:1969,Tatum:1971}.

The vibronic part $A^{e'v'}_{e''v''}$ of the transition
probability can be written as a function of it's vibronic
transition moment. This can be written in atomic units as

\begin{align}
\label{eq:Aevev}
A^{e'v'}_{e''v''}&=\frac{64\pi^{4}}{3hc^{3}}\nu^{3}_{v'v''}\frac{\left(2-\delta_{0,\Lambda'+\Lambda''}\right)}{\left(2-\delta_{0,\Lambda'}\right)}\left(\Re_{e}^{v'v''}\right)^{2}\\%
                 &=2.026\cdot10^{-6}\overline{\nu}^{3}\frac{\left(2-\delta_{0,\Lambda'+\Lambda''}\right)}{\left(2-\delta_{0,\Lambda'}\right)}\left(\Re_{e}^{v'v''}\right)^{2}\notag%
\end{align}

As the vibronic transition moment
$\left(\Re_{e}^{v'v''}\right)^{2}$ cannot be resolved for each
multiplet transition, a mean value of the transition
$\sum\left(\Re_{e}^{v'v''}\right)^{2}$ moment is rather used:

\begin{equation}
\left(\Re_{e}^{v'v''}\right)^{2}\cong\frac{\sum\left(\Re_{e}^{v'v''}\right)^{2}}
{(2-\delta_{0,\Lambda'+\Lambda''})(2S+1)}
\end{equation}

The vibronic transition moment can be calculated for each vibronic
transition as a function of the electronic transition moment and
the transition initial and final wavefunctions

\begin{equation}
\left(\Re_{e}^{v'v''}\right)^{2}=\left(\int\psi_{v'}(r)\Re_{e}(r)\psi_{v''}(r)dr\right)^{2}
\label{eq:Revv}
\end{equation}

Expressions for the electronic transition moment can be found in
the literature, either from spectroscopic measurements, or from
``ab-initio" calculations. This last method is usually preferred,
as nowadays, quantum methods have achieved a very good precision
\cite{Cramer:2004}. The vibrational wavefunctions are determined
by solving the radial Schr\"{o}dinger equation on the potential
curves of the upper and lower level potential curves.\bigskip

\begin{figure}[htb]
\begin{center}
\includegraphics[width=0.5\textwidth]{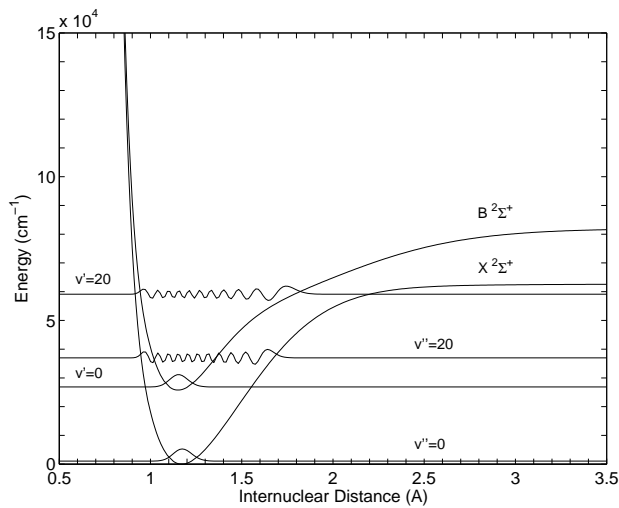}
\caption{\footnotesize{Example of an RKR calculation of the
potential curves of the upper and lower electronic states for the
CN Violet transition. The wavefunctions of different vibrational
levels of both electronic states are also reproduced at an
arbitrary scale}}
\label{fig:RKR}%
\end{center}
\end{figure}

Potential curves can be either calculated using ``ab-initio"
methods, or reconstructed through the Rydberg--Klein--Rees (RKR)
\cite{Rydberg:1931,Klein:1932,Rees:1947} method according to
experimental spectroscopic data. As modern spectroscopy is able to
resolve line positions to less than the \centi\rp\metre, level
energies can be known to a greater accuracy than using
``ab-initio" methods. However, ``ab-initio" methods are able to
reproduce the entire potential curve, whereas the RKR method can
only yield the region of the potential curve where measured data
is available. To overcome this problem, the central part of the
potential curve calculated through the RKR method is extrapolated
by a repulsive potential at narrower internuclear distances, and
by a Hulburth and Hirschfelder \cite{Hulburt:1941} potential at
larger internuclear distances, provided that the state
dissociation energy is know. This method can only be applied for
electronic states with a single potential with a shape close to a
well, but this is fortunately the case for most of the electronic
states for the molecules encountered in gas spectroscopy. A more
detailed overview of the calculation of potential curves and
vibronic wavefunctions can be found in \cite{LinodaSilva:2004-1}.
An example of a calculation of potential curves (using the RKR
method) and vibrational wavefunctions (solving the radial
Schr\"{o}dinger equation) is presented in Fig. \ref{fig:RKR}.

\subsection{Simulation of Linear Polyatomic Rovibrational Spectra Using the Line-by-Line Approach}

Calculations for linear polyatomic species present further
difficulties compared to diatomic species, as molecules have
several vibration modes (bending, symmetric/asymmetric stretch,
etc...). Emission spectra from these molecules results mainly from
rovibrational transitions, and a set of vibrational equilibrium
constants can still be defined, with a more complex formulation
(an example for the CO$_{2}$ molecule is presented in
\cite{Scutaru:1994}). However, estimation of these vibrational
parameters is rather difficult, the resulting values giving
inaccurate results in some cases. In practice, a set of
band-origin wavelengths and rotational constants for each
vibrational transition is given, allowing the calculation of the
line positions.\bigskip

The calculation of the lines intensities follows a similar
approach than for diatomic spectra calculations. Emission
coefficients of rovibrational transitions of polyatomic spectra
can be expressed through an expression similar to Eq.
\ref{eq:emicoeffdi}:

\begin{equation}
\varepsilon_{\nu}=\frac{1}{4\pi}N_{v'J'}A_{v'v''}S^{\ell'\phantom{'}J'}_{\ell''J''}F_{J'J''}h\nu%
\label{eq:emicoeffdirovib}
\end{equation}

Where the additional term $F_{J'J''}$ designates the Herman-Wallis
factor, which accounts for vibration-rotation interactions (see p.
110 in \cite{Herzberg:1965}). This factor has been omitted from
Eq. \ref{eq:emicoeffdi}, as vibration-rotation interactions can be
usually neglected in a rovibronic transition, owing to the usually
large energy gap between the transition electronic levels.
However, for rovibrational transitions, this interaction has to be
accounted for, which explains it's inclusion in Eq.
\ref{eq:emicoeffdirovib}.\bigskip

The vibrational Einstein coefficient is related to the vibrational
transition moment squared $\Re_{v'v''}$ through the relation
equivalent to Eq. \ref{eq:Aevev} for linear polyatomic
rovibrational transitions:

\begin{align}
A_{v'v''}&=\frac{64\pi^{4}}{3hc^{3}}\nu^{3}_{v'v''}\frac{(2-\delta_{0,l'})}{(2-\delta_{0,l''})}\Re_{v'v''}\\
&=2.026\cdot10^{-6}\bar{\nu}^{3}_{v'v''}\frac{(2-\delta_{0,l'})}{(2-\delta_{0,l''})}\Re_{v'v''}\notag
\end{align}

Determination of this squared transition moment $\Re_{v'v''}$ is
rather complex. Instead, values of the integrated intensity of a
vibrational band are tabulated at a reference temperature $T_{0}$
(usually 296 K) \cite{Taine:1983}. The value for the vibrational
dipole moment (in atomic units squared $(ea_{0})^{2}$) can then be
deduced from the following expression
\cite{Gamache:1992,Scutaru:1994}:

\begin{equation}
\Re_{v'v''}I_{a}=\frac{3hc}{8\pi^{3}}10^{43}\frac{S_{v'v''}^{0}}{\bar{\nu}_{v'v''}}
\frac{Q_{v0}}{(2-\delta_{0,l''})\exp\left(-\frac{hcE_{v'}}{k_{B}T_{0}}\right)}\frac{(ea_{0})^{2}}{D^{2}}%
\label{eq:RevvSvv}
\end{equation}

Expressions of the H\"{o}nl-London factors for parallel and
perpendicular transitions taken from \cite{Rothman:1992} are
presented in Tab. \ref{HLTri}.

\begin{table}[htb]
\begin{center}
\begin{tabular}{@{}c c@{}c@{}}
\toprule \midrule & $\Delta \ell=0$ & $\Delta \ell\neq 0$\\
\midrule P & $\frac{(J''+\ell'')(J''-\ell'')}{J''}$ & $\frac{(J''-1-\ell''\Delta \ell)(J''-\ell''\Delta \ell)}{2J''}$\\
Q & $\frac{(2J''+1)\ell''^{2}}{J''(J''+1)}$ & $\frac{(J''+1+\ell''\Delta \ell)(J''-\ell''\Delta \ell)(2J''+1)}{2J''(J''+1)}$\\
R & $\frac{(J''+1+\ell'')(J''+1-\ell'')}{J''+1}$ & $\frac{(J''+2+\ell''\Delta \ell)(J''+1+\ell''\Delta \ell)}{2(J''+1)}$\\
\bottomrule
\end{tabular}
\caption{\footnotesize{Honl-London Factors For Parallel and
Perpendicular
Vibrational Transitions}}%
\end{center}%
\label{HLTri}
\end{table}

Finally, the Herman-Wallis coefficients can be developed into the
following polynomial expansions:

\begin{center}
\begin{tabular}{@{}l l@{}}
P branch: & \!\!\!\!\!$(1-A_{1}J''+A_{2}J''^{2}-A_{3}J''^{3})^{2}$\\
Q branch: & \!\!\!\!\!$(1+A_{Q}J''(J''+1))^{2}$\\
R branch: & \!\!\!\!\!$(1+A_{1}(J''+1)+A_{2}(J''+1)^{2}+A_{3}(J''+1)^{3})^{2}$\\
\end{tabular}
\end{center}

values for the different coefficients $A$ being given for each
vibrational band.

\subsection{Line Shapes Calculations}

Line broadening mechanisms are grouped into two categories:
pressure broadening including natural, resonance, van der Waals,
and collisional broadening mechanisms. Pressure broadening is
represented by a Lorentzian profile for a given full width at half
maximum (FWHM) $\Delta\bar{\nu}_{L}$, and Doppler broadening is
represented by a Gaussian profile for a given FWHM
$\Delta\bar{\nu}_{G}$. A discussion on these different broadening
mechanisms can be found in \cite{Jefferies:1968}.\bigskip

The convolution of the line shapes resulting from these two types
of broadening mechanisms leads to a Voigt profile given by:

\begin{align}
v(\bar{\nu})=\frac{\Delta\bar{\nu}_{L}}{\Delta\bar{\nu}_{G}}\sqrt{\frac{\ln2}{\pi^{3}}}\int^{+\infty}_{-\infty}
\frac{\exp\left\{-\frac{\left[(\xi-\bar{\nu}-\bar{\nu}_{0})^{2}\ln2\right]}{\Delta\bar{\nu}_{G}^{2}}\right\}}{\xi^2+\Delta\bar{\nu}_{L}^{2}}d\xi
\end{align}

However, the explicit convolution of these two line shapes can
lead to high calculation times, and semi-empirical expressions
approaching this exact line profile are seldom used.\bigskip

An empirical expression for this Voigt profile is given by Arnold
\cite{Arnold:1969}\footnote{the term 4$\ln2$ can be replaced by
2.772 for computational efficiency}:

\begin{align}
v(\bar{\nu})&=C_{1}e^{-4\ln2D^{2}}+\frac{C_{2}}{1+4D^{2}}\cdots\notag\\
&+0.016C_{2}\left(1-\frac{\Delta\bar{\nu}_{L}}{V}\right)\left(e^{-0.4D^{2.25}}-\frac{10}{10+D^{2.25}}\right)
\end{align}

where
\begin{align*}
D&=\frac{\bar{\nu}-\bar{\nu}_{0}}{V}\\
V&=\frac{1}{2}\left(\Delta\bar{\nu}_{L}+\sqrt{\Delta\bar{\nu}_{L}^{2}+4\Delta\bar{\nu}_{G}^{2}}\right)\\
C_{1}&=\frac{\left(1-\frac{\Delta\bar{\nu}_{L}}{V}\right)}{V\left(1.065+0.047\frac{\Delta\bar{\nu}_{L}}{V}+0.058\frac{\Delta\bar{\nu}_{L}}{V^{2}}\right)}\\
C_{2}&=\frac{\left(\frac{\Delta\bar{\nu}_{L}}{V}\right)}{V\left(1.065+0.047\frac{\Delta\bar{\nu}_{L}}{V}+0.058\frac{\Delta\bar{\nu}_{L}}{V^{2}}\right)}\\
\end{align*}

The accuracy to the exact expression being within 1 \%.\bigskip

A simple modification to the Voigt FWHM expression has been
proposed by Olivero \cite{Olivero:1977}:

\begin{align*}
V&=\frac{1}{2}\left(1.0692\Delta\bar{\nu}_{L}+\sqrt{0.86639\Delta\bar{\nu}_{L}^{2}+4\Delta\bar{\nu}_{G}^{2}}\right)\\
\end{align*}

giving a better accuracy of 0.02 \% to the exact
expression.\bigskip

This presented lineshape calculation method presents the advantage
of allowing a good accuracy to the exact function, without a great
computational burden. However, many other methods exist for the
calculation of a Voigt lineshape. An overview and discussion for
these different methods can be found in \cite{Schreier:1992}.

\subsection{Strategies for Fast and Accurate Spectral Simulations}

Spectral calculations using the line-by-line approach can lead to
large calculation times, as several thousands of lines are
typically calculated and convoluted with the corresponding
lineshapes. Also, such calculations can return spectral grids
which can be several million points wide. Therefore, some
techniques may be used in order to reduce the number of calculated
lines. A good way to achieve this consists in making some
approximations regarding the fine-structure (spin-splitting
effects) of the simulated spectra. This fine structure can in some
cases be entirely or partially neglected resulting in lesser
computed lines.

\subsubsection{neglecting line spin-splitting effects}%
\label{negspin}

Taking into account the effects of spin-splitting in calculations
of the fine structure for multiplet transitions leads to the
determination of up to 12 and 27 rotational branches for doublet
and triplet transitions respectively instead of the 3 usual $P,Q$
and $R$ branches. This results in calculation times increased by a
factor of 4 and 9 respectively. Therefore, one may want to neglect
spin-splitting in order to reduce computation times, but without
losing precision in the calculation result. One needs then to
evaluate the importance of line spin-splitting over the line
resolution (which depends on the lines FWHM). If the separation of
multiplet lines is much smaller than their widths, such lines will
be accurately modelled by one singlet line.\bigskip

This approach is usually valid for multiplet $\Sigma$ states,
which have small spin-splitting factors allowing transitions
between such states to be accurately modelled as singlet states.

A good example can be given for a $^{2}\Sigma-^{2}\Sigma$
transition. For this case, the separation between two doublet
lines (in \AA) is given by:

\begin{equation}
\Delta\lambda\simeq10^{8}\frac{(\gamma_{E}-\gamma_{G})(J+\frac{1}{2})}{\bar{\nu}_{0}^{2}}
\end{equation}

For the CN violet system\footnote{the spin-spin correction is
$\gamma=7.26\cdot 10^{-3}$cm$^{-1}$ for the (X$^{2}\Sigma$,v=0)
state, and $\gamma=17.16\cdot 10^{-3}$cm$^{-1}$ for the
(B$^{2}\Sigma$,v=0) state \cite{Prasad:1992}}, this is equivalent
to a line splitting of 0.14 \AA\ for a rotational value as high as
J=100. In classical spectrometry applications, this transition can
be accurately simulated as a singlet transition, reducing
calculation times without a loss of accuracy. For triplet $\Sigma$
transitions, a slightly more complicated relationship can also be
determined to evaluate whether the analyzed transitions can be
accurately modelled as singlet transitions (according to the
transition $\gamma$ and $\lambda$ constants).\bigskip

The limits of this assumption occur when the spin-splitting of the
studied states is no more negligible, and self-absorption of the
spectra is high. In this case, a comparison of a simulated
spectrum using this simplification with a simulated spectrum
accounting for spin-splitting shows non-negligible differences
\cite{Arnold:1969}. However, even for multiplet states with larger
spin-splitting values such as $\Pi$ and $\Delta$, the same
approach may still be valid for lower resolution spectra (with
line widths of about 1--10 \AA\ or higher).

\subsubsection{Reducing the number of simulated rotational branches}

Even when spin-splitting effects need to be accounted for in
spectral simulations, one may still reduce the number of
calculated rotational branches making use of certain
simplifications.\bigskip

The first case occurs for perpendicular transitions between
$\Sigma$ and $\Pi$ states when neglecting the spin-splitting of
the $\Sigma$ state using the approach described in section
(\ref{negspin}) leads to the superposition of several branches.

As an example, for $^{2}\Sigma\leftrightarrow^{2}\Pi$ doublet and
for $^{3}\Sigma\leftrightarrow^{3}\Pi$ triplet transitions 4 and
10 satellite branches respectively coalesce with the main branches
when neglecting spin-splitting of the $\Sigma$ state.\bigskip

The second case occurs for calculations of high temperature
spectra in applications such as combustion or plasma radiation in
which some branches have small intensities and can be neglected.

Energy exchanges between the different rotational states (R-R
transfer) are very efficient which means that in most of the
spectroscopic applications of interest, a Boltzmann distribution
of these states may be assumed. Therefore, a line rotational
intensity may be written as

\begin{equation}
S(J)=\frac{(2J+1)\exp\left(-\frac{hc}{kT_{rot}}\,F(J)\right)}{Q_{rot}}\frac{S^{\Lambda'\phantom{'}J'}_{\Lambda''J''}}{2J+1}
\end{equation}

as each rotational state has a $g_{J}=2J+1$ degeneracy, the ground
state will not be the most populated state. As an example, at room
temperature, typically the most populated rotational level is
around $J=7-10$, whereas at a 2000 \kelvin\ temperature, the most
populated level is around $J=20$.

\begin{figure}[htb]
\begin{center}
\includegraphics[width=0.5\textwidth]{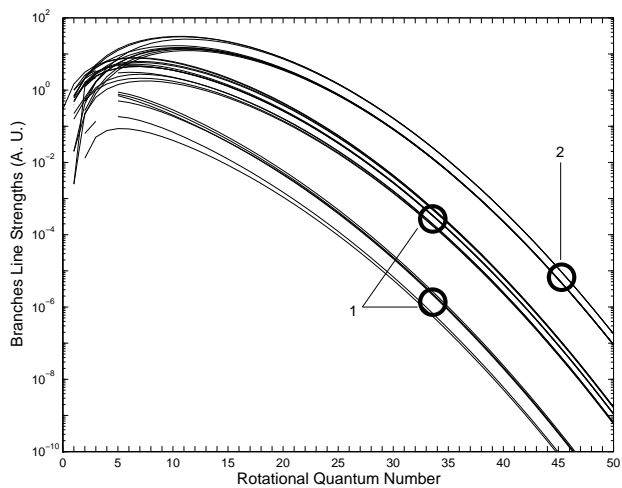}
\caption{\footnotesize{Rotational line strengths of the different
branches of a $^{3}\Sigma\leftrightarrow^{3}\Pi$ transition for a
Boltzmann distribution of the rotational levels at a
characteristic temperature of 300 \kelvin. Expressions for the
H\"{o}nl-London factors are taken from \cite{Budo:1937}. The
corresponding weak branches (group 1) are listed in appendix
\ref{sec:HLBranches}}}
\label{fig:HL3S3P}%
\end{center}
\end{figure}

This is an advantage for the simulation of high temperature gases
radiation as the rotational transition probabilities (Honl-London
factors) for some branches have smaller magnitudes than others,
only reaching equivalent magnitudes for the first rotational
levels (typically $J\leq5$). If we take into account the fact that
those levels are not the most populated according to what was said
before, it can be verified that the peak rotational line strengths
of such branches are typically more than one order of magnitude
smaller than the peak rotational strengths of other branches. An
example for a $^{3}\Sigma\leftrightarrow^{3}\Pi$ transition at a
300 \kelvin\ gas temperature is presented in Fig. \ref{fig:HL3S3P}

The ratio of the peak intensities between the second and first
group of rotational branches is 3.3 for this case. For a 2000
\kelvin\ temperature, the same calculation results in a 15.8
ratio. Therefore omitting such weaker rotational branches for
spectral calculations of high temperature gases has in most cases
a negligible effect on the simulated spectra.\bigskip

Branches than can be neglected and coalesced branches when the
spin-splitting of $\Sigma$ states in perpendicular transitions is
neglected are presented in appendix \ref{sec:HLBranches} for
doublet and triplet transitions between $\Sigma$ and $\Pi$ states.

\section{Examples of Molecular Radiation Simulation in the Line-by-Line Approach}

Some applications of the different methods for the calculation of
line positions are presented in this section. Specific spectra is
numerically reproduced, and the applicability of each of these
methods is discussed. The experimental data presented here has
been recorded at high resolution 
in low pressure plasma facilities, using emission spectroscopy
techniques.

\subsection{Experimental Setup}

The molecular spectra presented in this study is issued from two
similar low-pressure facilities where strongly emissive plasmas
are obtained. The two facilities are located at the CORIA
Laboratory in Rouen and the Laboratoire d'A\'{e}rothermique in
Orl\'{e}ans respectively, both laboratories from the Centre
National de la Recherche Scientifique in France. Specific
molecular systems were acquired using emission spectroscopy
techniques, with typical resolutions of 0.3--0.5 \angstrom.

\subsubsection{The SR5 Arc-jet Plasma Wind-tunnel}

The SR5 low pressure arc-jet plasma wind-tunnel available at the
Laboratoire d'A\'{e}rothermique has been used during the last
years for the simulation of the entry conditions in Earth, Mars,
and Titan planetary atmospheres \cite{Lago:2001}.

A D.C. vortex-stabilized arc operating at low voltages (50--100
\volt) and low currents (50--150 \ampere) delivers typical powers
of 5--10 \kilo\watt\ to the flow in the throat region of the
nozzle. The low mass-flow rates (0.1--0.5 \gram\per\second)
crossing the nozzle allow the obtention of an high-enthalpy (5--30
\mega\joule\per\kilogram) steady plasma jet with a global yield of
50-70\%. This level of specific enthalpy obtained with a small
mass-flow rate is an advantage because of the low electrode
erosion, which allows maintaining a steady plasma jet for several
hours with a low level of contamination. The pumping system
capacity of 26.000 \cubic\metre\usk\rc\hour\ ensures an ambient
pressure of about 10 \pascal\ to be maintained in a 4.3 \metre\
long and 1.1 \metre\ diameter vacuum chamber in which the arc jet
is expanded.

Spectral measurements were carried using a SOPRA F1500
(Ebert-Fastie type) monochromator with a focal length of 1500
\milli\meter\ and a grating of 1800 grooves/\milli\metre\ sweeping
a spectral region from 270 \nano\metre\ in the near-ultraviolet
region to 950 \nano\metre\ in the near-infrared region. The
grating is connected to an intensified optical multichannel
analyser (Princeton Instruments IRY 1024). This device allows a
8.5-\nano\metre\ wavelength region to be expanded on 1024 pixels
and is cooled by a Peltier element insuring an operating
temperature of -35\celsius. The plasma is imaged onto the
monochromator by a mirror telescope connected to the entrance slit
by a quartz optical fiber. The entrance slit opening can be
adjusted, therefore modifying the experimental apparatus function
which can reach values down to a full width at half maximum (FWHM)
of 0.03 \nano\meter.

\subsubsection{The CORIA Inductively Coupled Plasma Torch}

The CORIA inductively coupled plasma (ICP) torch is used for
different research fields such as plasma de-pollution applications
as well as the investigation of plasma radiation from the vacuum
ultraviolet to the infrared regions \cite{Thebault:1998}.

The facility includes a high frequency (1.76 \mega\hertz)
generator, with typical voltages and intensities around 7
\kilo\volt\ and 10 \ampere\ respectively, producing an
electromagnetic field onto a five fingers coil surrounding a
quartz chamber in which a test gas is inserted. The typical mass
flow rates of 1--5 \gram\per\second, for a delivered power of
about 70 \kilo\watt\per\second\ and a global yield of around 30 \%
lead to the formation of a high enthalpy (around 10
\mega\joule\per\kilogram) stable and homogeneous plasma. Further,
a convergent nozzle can be added between the torch and a 1 \metre\
long and 0.5 \metre\ diameter test-chamber in which measurements
can be implemented.

UV and visible spectra of air plasmas is acquired using an Acton
Research spectrometer of 750 \milli\metre\ focal length
implemented with a Princeton IMAX Intensified camera allowing
spectral resolutions up to 0.045 \nano\metre.

\subsection{Presented Examples}

An investigation has been carried on the numerical reproduction of
molecular spectra from the CN Violet System $\Delta v=0$, the
C$_{2}$ Swan Bands $\Delta v=0$, and the N$_{2}^{+}$ First
Negative System $\Delta v=0$. Each of these systems have been
investigated previously \cite{LinodaSilva:2004-1,LinodaSilva:2003}
and difficulties linked to the numerical description of such
spectra have motivated their selection for a discussion in this
work.

\subsubsection{The CN Violet System}

Conducting spectral measurements of line positions in low pressure
and temperature gases and plasmas allows a precise and unambiguous
line position determination, as broadening mechanisms (principally
doppler broadening, which is temperature dependent) will be kept
to a strict minimum. However, compared to higher temperature
measurements, only the lower energy vibronic states will be
excited, and less lines will be measured. Therefore, the
interpolation of such line positions through the usual polynomial
expressions will only be exactly valid for the lower vibrational
and rotational levels. Extrapolation of such polynomial expansions
to higher levels will of course be risky, specially taking into
account the strong oscillations of the higher terms of such
expansions. Consequently, a good balance between the accuracy of
the line positions determination and the number of measured lines
has to be achieved.

The CN Violet system is a strong radiative emitter in most air and
carbon species plasmas. Therefore, it has been extensively studied
by different research teams for over 50 years, and various sets of
spectroscopic constants have been proposed for the calculation of
line positions. Such sets have been obtained through line position
measurements to different levels of precision, and to different
temperature ranges. Namely, fourier spectroscopy techniques have
been applied to the measurement of this radiative system over the
more recent years.\bigskip

Emission from the CN Violet system $\Delta v=0$ has been measured
in the SR5 facility for a low pressure and high enthalpy
N$_{2}$--CH$_{4}$ plasma, in which a Boltzmann equilibrium of the
CN molecule rotational and vibrational levels is achieved. The
apparatus function of the measurement system has been held at a
FWHM of 1 \angstrom, so as to resolve solely the vibrational bands
of the system and to allow an easier comparison of the different
spectroscopic sets available regarding vibrational bandheads.

\begin{figure}[!t]
\begin{center}
\includegraphics[width=0.5\textwidth]{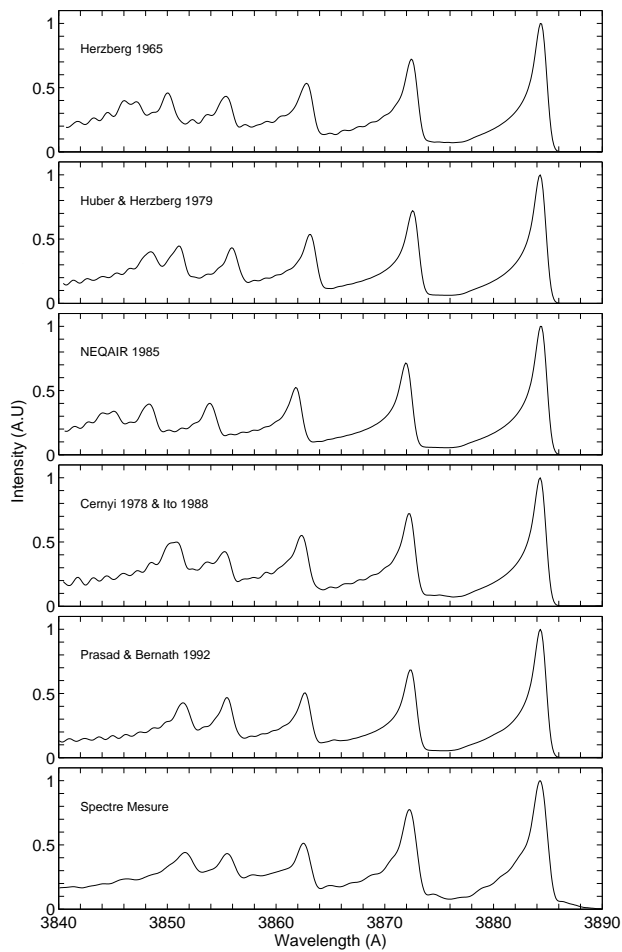}
\caption{\footnotesize{Comparison between a CN violet system
$\Delta$v=0 spectrum measured in a CH$_{4}$--N$_{2}$ (0.002
\gram\per\second, 0.36 \gram\per\second, 6.7 \kilo\watt\ total
power, 4 \kilo\watt\ injected power, 3.8 \pascal\ background
pressure) plasma in the SR5 facility and simulated spectra using
different spectroscopic constants. The molecule vibrational and
rotational temperatures were iteratively set in the calculation
until a best fit was achieved. The spectrometer apparatus function
was determined using a mercury calibration lamp. The different
parameters read:
FWHM=1\,\AA, T$_{v}\!=\!10800\pm100\,\kelvin$,%
T$_{r}\!=\!4900\pm100\,\kelvin$}}%
\label{fig:CN}
\end{center}
\end{figure}

The spectra has been numerically reproduced using the set of
spectroscopic constants proposed by Herzberg \cite{Herzberg:1965},
the set of spectroscopic constants proposed by Huber \& Herzberg
\cite{Huber:1979}, the set of spectroscopic constants used in the
1985 version of the well known line-by-line spectral code NEQAIR
\cite{NEQAIR:1985}, the sets of spectroscopic constants proposed
by Cerny \cite{Cerny:1978} for the X$^{2}\Sigma^{+}$ level and Ito
\cite{Ito:1988} for the B$^{2}\Sigma^{+}$ level, and finally the
set of spectroscopic constants proposed by Prasad \& Bernath
\cite{Prasad:1992}. The transition probabilities have been taken
from the set proposed by Knowles \cite{Knowles:1988}. A comparison
between the measured spectrum and the reproduced spectra, using
the above constant sets and a line shape simulating the
experimental FWHM, is presented in Fig. \ref{fig:CN}.

Large discrepancies can easily be detected between the different
simulated spectra. All the spectroscopic constants reproduce to a
good level of accuracy the 0--0, 1--1, and 2--2 bandheads of the
experimental spectrum. However, for higher vibrational bands,
discrepancies tend to become larger, and some simulated spectra
fails to reproduce accurately the band reversal observed in the
experimental spectrum which leads to the superposition of the 4--4
and 5--5 bandheads. Older spectroscopic constant sets such as the
ones proposed by Herzberg and Huber \& Herzberg are inadequate as
they likely did not take into account higher vibrational levels
when they were developed, resulting in a limited validity range.

The poor reproducibility of the measured spectrum by the simulated
spectrum using the spectroscopic constants set of the NEQAIR85
code is likely due to the truncation to the value of
$\omega_{e}y_{e}$ for the vibrational levels polynomial expansion
(see Eq. \ref{eq:KleinDunham1}). Another example of such issues is
discussed in \cite{Laux:1993} regarding the simulation of the
N$_{2}$ 2nd Positive System. Care should therefore be exercised
when using spectroscopic sets with large polynomial expansions in
numerical codes which do not account for higher order terms.
Resorting to the truncation of available sets may still yield
correct approximations for the lower vibrational levels energies,
but for higher levels energies, which are more dependant on higher
order terms of such polynomial expansions, this is no longer the
case. Instead, it is preferable to fit the initial polynomial
expansion to a lower order polynomial expansion, using the
validity range of the initial fit for the level energies
calculations.

More recent spectroscopic datasets, determined through spectral
measurements using Fourier spectroscopy techniques, reproduce more
accurately the experimental spectrum. The spectroscopic constants
proposed by Cernyi and Ito already allow reproducing correctly the
4--4 and 5--5 bandhead superposition. However, the 6--6 bandhead
still does not match accurately the experimental spectrum.
Finally, the most recent spectroscopic constants, proposed by
Prasad allow an exact reproduction of the experimental spectrum
(to the measured resolution).\bigskip

\subsubsection{The C$_{2}$ Swan Bands}

Various sets of equilibrium spectroscopic constants are proposed
by different authors for the simulation of different diatomic
spectra. For most applications, the accuracy of such datasets is
sufficient for carrying spectral simulations, provided that an
analysis of the available data is carried and the more accurate
data is selected. However, in some minor cases equilibrium
spectroscopic constants no longer suffice for a correct
description of measured spectra. The frequently observed Swan
Bands of the C$_{2}$ molecule provide a good example for when
equilibrium spectroscopic constants fail to accurately reproduce
experimental spectra.\bigskip

\begin{figure}[!t]
\begin{center}
\includegraphics[width=0.5\textwidth]{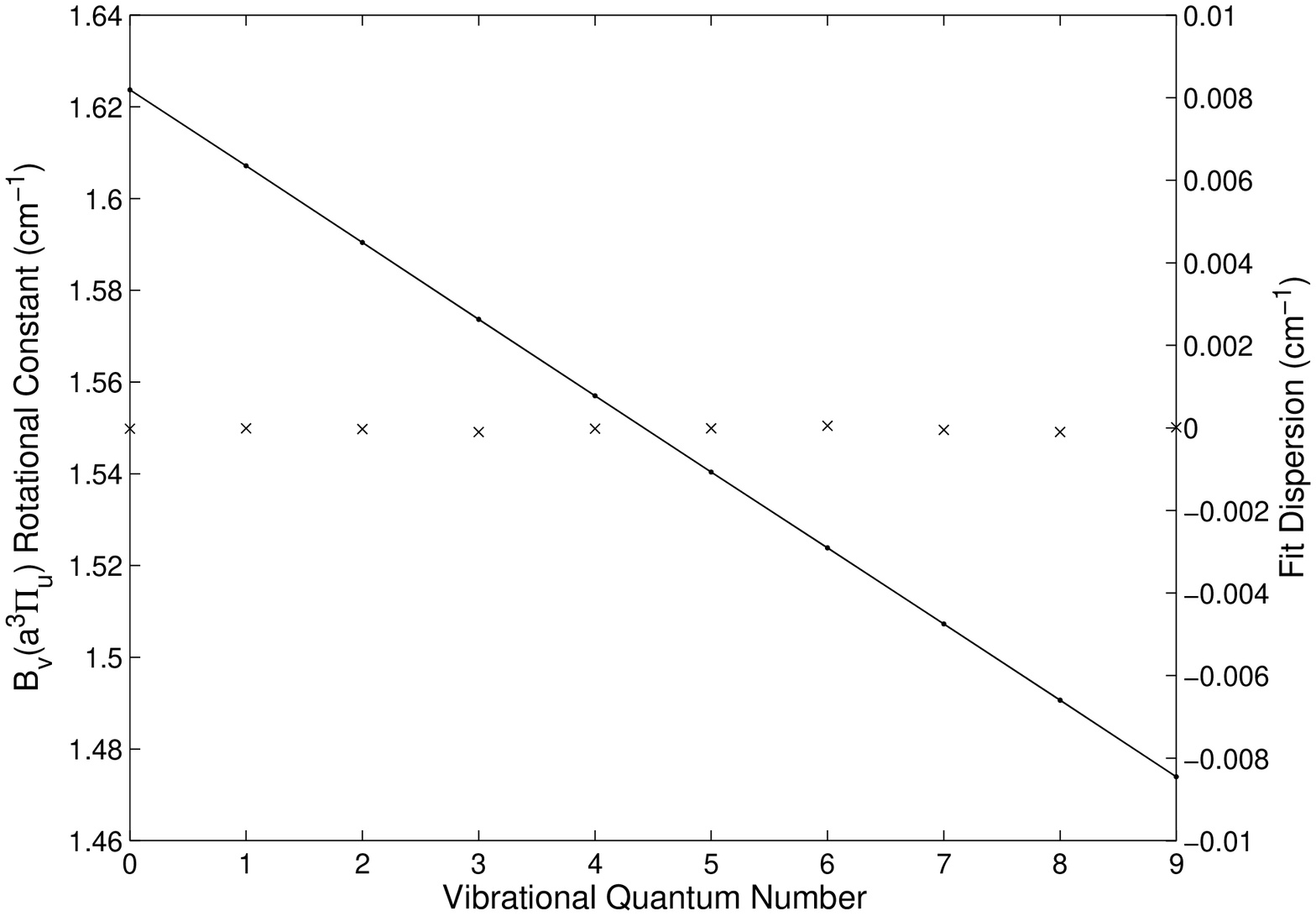}
\includegraphics[width=0.5\textwidth]{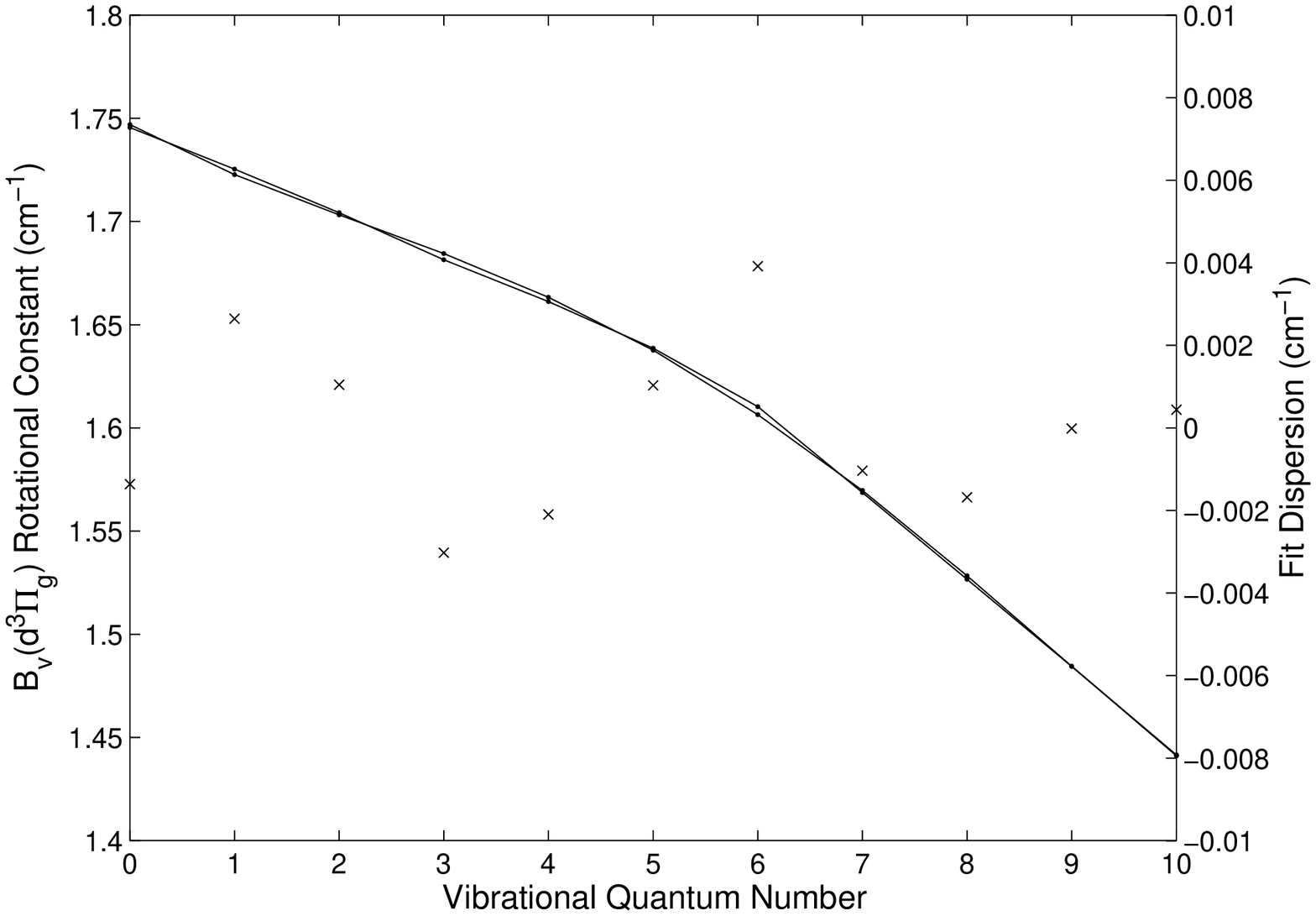}
\caption{\footnotesize{Comparison between level and interpolated
rotational constants B$_{v}$ for the a$^{3}\Pi_{u}$ and the
d$^{3}\Pi_{g}$ electronic states of C$_{2}$. A vibrational
perturbation of the d$^{3}\Pi_{g}$ state leads to a trend change
for the rotational constant above the v=5 level, preventing an
accurate fit to the usual equilibrium spectroscopic constants from
being obtained}}%
\label{fig:C2fit}
\end{center}
\end{figure}

A good starting point for the analysis of the spectroscopic
constants for the upper and lower electronic states of this system
is given by a study from Phillips \cite{Phillips:1968}, which
analyzed the first 10 vibrational levels of such states. Although
the vibrational levels of the a$^{3}\Pi_{u}$ state follow the
usual monotonous evolution pattern typical of Eqs.
\ref{eq:KleinDunham1}, \ref{eq:KleinDunham2} and
\ref{eq:KleinDunham3}, the vibrational levels of the
d$^{3}\Pi_{g}$ state show a consistent change of trend, starting
from the vibrational level $v$=5. In this specific case, it is no
longer possible to fit the level spectroscopic constants $G_{v}$,
$B_{v}$, $D_{v}$, $\ldots$ to the usual polynomial expansions. A
fit of the rotational constants $B_{v}$ provided by Phillips to
the fifth order polynomial expansion of Eq. \ref{eq:KleinDunham2}
has been carried for both the electronic levels, and the
interpolated values for $B_{v}$ are compared to the initial values
in Fig. \ref{fig:C2fit}.

It can be easily verified that, although the values of $B_{v}$ for
the a$^{3}\Pi_{u}$ electronic state can be exactly reproduced by a
polynomial expansion, this is no longer the case for the
d$^{3}\Pi_{g}$ electronic state where the mathematical expression
of Eq. \ref{eq:KleinDunham2} is no longer adapted to the
description of the level-dependent rotational constants $B_{v}$,
yielding a large fit dispersion.\bigskip

\begin{figure}[htb]
\begin{center}
\includegraphics[width=0.5\textwidth]{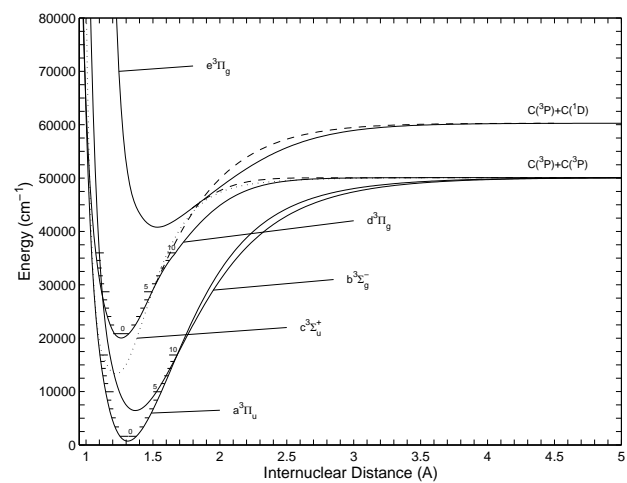}
\caption{\footnotesize{Potential curves of the d$^{3}\Pi_{g}$,
a$^{3}\Pi_{u}$ and neighboring triplet electronic states obtained
using the RKR method. The dashed lines represent the shape of the
d$^{3}\Pi_{g}$ and e$^{3}\Pi_{g}$ states potential curves if the
vibrational perturbation resulting from the avoided intersection
of the potential curves is not accounted for}}%
\label{fig:C2pot}
\end{center}
\end{figure}

The origin of the perturbation observed for the spectroscopic
constants of the d$^{3}\Pi_{g}$ electronic state, starting from
level $v$=5 has been investigated through an analysis of the
reconstructed potential curves of the transition and the
neighboring triplet states, using the RKR method. Spectroscopic
constants for the b$^{3}\Sigma^{-}_{g}$ and e$^{3}\Pi_{g}$ states
were taken from the compilation of Huber \& Herzberg
\cite{Huber:1979}, and the spectroscopic constants for the
c$^{3}\Sigma^{+}_{u}$ state were taken from Ballik and Ramsay
\cite{Ballik:1963} who determined them through the study of
perturbations in the A$^{1}\Pi_{u}$ state. The calculated
potential curves are presented in Fig. \ref{fig:C2pot}.

An analysis of the curve plots show that the perturbation of the
d$^{3}\Pi_{g}$ electronic state could be explained by the crossing
of the b$^{3}\Sigma^{-}_{g}$ potential curve which occurs at
$v$=5. However for such cases, the perturbation would remain
local, and higher vibrational levels of the d$^{3}\Pi_{g}$ state
would remain unperturbed, which is not the case. Therefore, the
more probable cause for this perturbation could result from an
avoided intersection between the d$^{3}\Pi_{g}$ and e$^{3}\Pi_{g}$
potential curves (see p. 295 in \cite{Herzberg:1965}), which
alters the curve shape above the perturbation.

Dashed curve lines have been added to the plot to show the avoided
intersection between the potential curves. The reconstruction of
the unperturbed potential curve for the d$^{3}\Pi_{g}$ electronic
state has proceeded including solely the unperturbed levels
$v$=0--4 in the RKR calculation. The extrapolation of the
potential curve by a Hulburth and Hirschfilder expression has
taken into account a higher dissociation level
[C($^{3}$P)+C($^{1}$D)]. The reconstruction of the e$^{3}\Pi_{g}$
electronic state potential curve has proceeded identically, taking
into account a lower dissociation level
[C($^{3}$P)+C($^{3}$P)].\bigskip

\begin{figure}[htb]
\begin{center}
\includegraphics[width=0.5\textwidth]{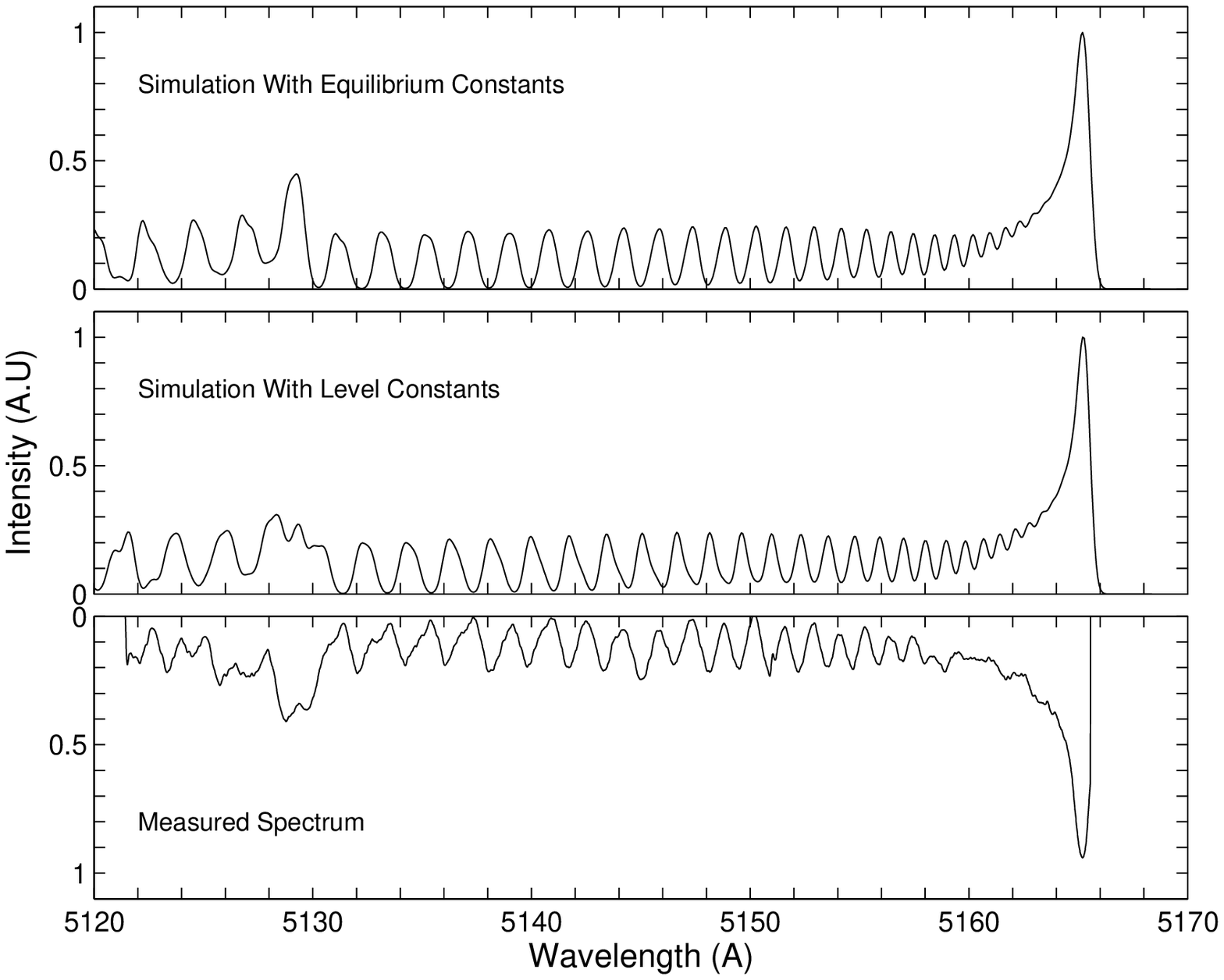}
\caption{\footnotesize{Comparison between a C$_{2}$ Swan bands
$\Delta$v=0 spectrum measured in an CO$_{2}$--N$_{2}$--Ar (2
\gram\per\second, 0.2 \gram\per\second, 0.2 \gram\per\second, 70
\kilo\watt\ total power, 20 \kilo\watt\ injected power, 1000
\pascal\ background pressure, no nozzle) plasma in the
\hbox{CORIA} ICP torch using equilibrium and level spectroscopic
constants. The molecule vibrational and rotational temperatures
were iteratively set in the calculation until a best fit was
achieved. The spectrometer apparatus function was determined using
a mercury calibration lamp. The different parameters read:
FWHM=0.55\,\AA, T$_{v}$=T$_{r}\!=\!4200\pm100\,\kelvin$}}%
\label{fig:C2}
\end{center}
\end{figure}

An experimental spectrum from the C$_{2}$ Swan Bands $\Delta$v=0,
obtained in the CORIA ICP torch for a low pressure
CO$_{2}$--N$_{2}$--Ar plasma flow, has been simulated using the
level constants proposed by Phillips, and their interpolation to
equilibrium constants. The transition probabilities used in the
calculation are issued from Cooper \cite{Cooper:1976}. The
comparison of the simulated spectra, using level and equilibrium
spectroscopic constants to the experimental spectrum is presented
in Fig. \ref{fig:C2}.

As it could be expected after the analysis of the fits presented
in Fig. \ref{fig:C2fit}, it is verified that equilibrium
constants, although fitted to very accurate level constants, fail
to reproduce exactly the experimental spectrum. This is
particularly apparent for the 1--1 bandhead at 5120 \angstrom,
which is not very well reproduced unlike the spectrum issued from
level spectroscopic constants. Therefore, whenever vibrational
perturbations are present for any of the states of the radiative
transition, it is likely that the explicit insertion of level
spectroscopic constants in simulation codes will be a required
feature for a sufficiently accurate simulation.

\subsubsection{The N$_{2}^{+}$ First Negative System}

A final example can be presented for a spectrum affected by
rotational perturbations of it's internal states (intersection of
potential curves for $J\neq0$). For this case, there is a shift of
the nearby levels populations following a mathematical expression
of the type $1/x$ (see p. 283 in \cite{Herzberg:1965}). The more
accurate but more complex method for accounting for the presence
of such perturbations consists in solving the perturbative
hamiltonian for the system \cite{Lefebvre-Brion:1986}. However, in
our case, it was chosen to resort to the usual Klein--Dunham
coefficients for the calculation of the level energies, with the
addition of a term in the form of

\begin{equation}
E_{J}=E_{J}+\frac{\Delta E_{max}}{2\left(J-J_{pert}-1/2\right)}
\end{equation}

The upper electronic state B$^{2}\Sigma^{+}_{u}$ of the
N$_{2}^{+}$ First Negative System is known to be perturbed by the
A$^{2}\Pi_{u}$ electronic state. The influence of the shifts of
levels energies on the observed spectrum are quickly appreciable,
which makes this radiative system a good candidate for the
illustration of the importance of perturbations modelling in
spectral simulations. Such perturbations have been extensively
studied in \cite{Michaud:2000,Michaud:1996}, and the values for
$J_{pert}$ and $\Delta E_{max}$ (position and intensity of the
perturbation) for the perturbations resulting from the above
described interaction are reported in the references therein.

\begin{figure}[htb]
\begin{center}
\includegraphics[width=0.5\textwidth]{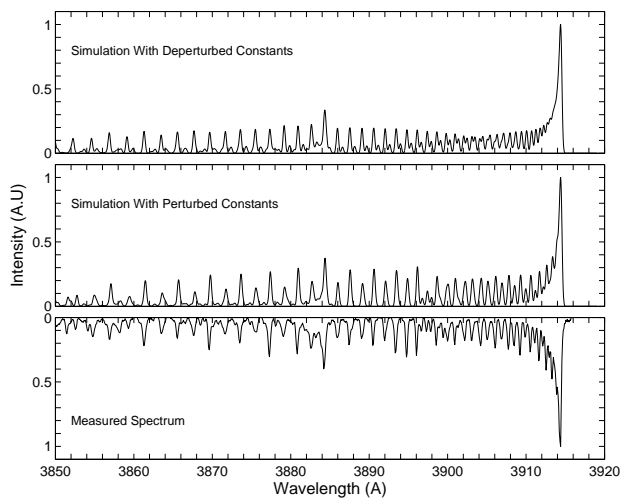}
\caption{\footnotesize{Comparison between a N$_{2}^{+}$ 1st
negative system $\Delta$v=0 spectrum measured in an air (0.2
\gram\per\second, 4.6 \kilo\watt\ total power, 2.4 \kilo\watt\
injected power, 4.4 \pascal\ background pressure) plasma in the
SR5 facility and simulated spectra with and without the simulation
of perturbations. The molecule vibrational and rotational
temperatures were iteratively set in the calculation until a best
fit was achieved. The spectrometer apparatus function was
determined using a mercury calibration lamp. The different
parameters read: FWHM=0.33\,\AA, T$_{v}\!=\!4000\pm100\,\kelvin$,
T$_{r}\!=\!2600\pm100\,\kelvin$}}%
\label{fig:N2II}
\end{center}
\end{figure}

Emission from the N$_{2}^{+}$ First Negative System system $\Delta
v=0$ has been measured in the SR5 facility for a low pressure air
plasma. The apparatus function of the measurement system has been
set at a FWHM of 0.3 \angstrom, which allows fully resolving the
rotational lines of the spectrum. Spectral simulations have then
been carried for this molecular system, firstly excluding, and
lately taking into account the influence of the perturbations in
the spectral model. Level spectroscopic constants provided by
Michaud \cite{Michaud:2000} and transition probabilities provided
by Laux \cite{Laux:1992} have been taken into account for the
calculations. The obtained results and a comparison with the
experimental spectrum are presented in Fig. \ref{fig:N2II}.

The analysis of the simulated spectra allows detecting large
discrepancies due to the presence of perturbations. Namely, the
structure of the 0--0 vibrational band is completely altered by
the perturbations of the upper levels energies. The effects of the
perturbations in the simulated spectrum are very apparent near
3900 \angstrom. It can be verified that the simplified model used
for including the effects of perturbations in the level energies
allows obtaining a very close match to the experimental spectrum.
This is no longer the case for the unperturbed spectrum, which
fails to reproduce correctly the experimental spectrum, although
accurate level spectroscopic constants are used in the
calculation.

\section{Conclusion}

The methods and issues regarding line-by-line spectral
calculations have been discussed in this paper. Accurate methods
for calculating the main parameters of bound radiation such as
line positions, intensities, and shapes, have been presented for
diatomic rovibronic transitions, and linear polyatomic
rovibratinal transitions.

Moreover, as line-by-line molecular simulations lead to the
calculation of spectrums composed from a very large number of
lines, with arising issues regarding computation times and memory
limitations, strategies leading to a reduction in the number of
calculated lines, with minor accuracy losses, have been presented.

A more complete discussion on the different methods available for
the calculation of line positions has also been carried.
Additionally to the necessity of selecting the most adapted method
for the calculation of line positions, the importance of selecting
the most accurate and adapted spectral dataset has been
highlighted. It has been showed that coping with the specific
issues of the simulation of unperturbed but also perturbed spectra
may allow reaching simulated spectra that matches very accurately
experimentally determined spectra. Moreover, achieving such
accuracies in spectral simulations also results in a very suitable
method for determining vibrational and rotational temperatures
$T_{v}$ and $T_{r}$ to a good level of accuracy. The methods which
were applied to the simulation of the presented experimental
spectra have allowed determining the vibrational and rotational
temperatures to an accuracy of 100 \kelvin. Even when a Boltzmann
equilibrium is not achieved for such levels, such methods can be
successfully applied to the determination of the levels population
distributions \cite{LinodaSilva:2004-2}.

\section*{\small{Acknowledgements}}

The author would like to acknowledge P. Boubert for providing the
experimental spectrum of the C$_{2}$ Swan Bands presented in this
paper.

\appendix

\section*{Appendix}

\section{Multiplet Level Energies}
\label{sec:LevEn}

All the expressions below are written in wavenumber units
(\centi\rc\metre).

\subsection*{Doublet Levels}

The general expression for the doublet energy levels is given by
\cite{Herzberg:1965}:

\begin{subequations}
\begin{align}
F_{3/2}(J\!\geq\!1)&=B_{v}
\left[%
\begin{array}{l}%
\left(J+\frac{1}{2}\right)^{2}
-\Lambda^{2}\\
-\frac{1}{2}\left(4\left(J+\frac{1}{2}\right)^{2}
+Y(Y-4)\Lambda^{2}\right)^{\frac{1}{2}}
\end{array}%
\right]\notag\\
&-D_{v}J^{4}\\
F_{1/2}(J\!\geq\!0)&=B_{v}\left[%
\begin{array}{l}%
\left(J+\frac{1}{2}\right)^{2}
-\Lambda^{2}\\
+\frac{1}{2}\left(4\left(J+\frac{1}{2}\right)^{2}
+Y(Y-4)\Lambda^{2}\right)^{\frac{1}{2}}
\end{array}%
\right]\notag\\
&-D_{v}(J+1)^{4}
\end{align}
\end{subequations}

For $^{2}\Sigma$ states, a simpler expression of the levels
energies is used \cite{Herzberg:1965}:

\begin{subequations}
\begin{align}
^{2}\Sigma_{3/2}(J\!\geq\!1)&=B_{v}(J(J+1))-D_{v}(J(J+1))^{2}+\gamma\left(\frac{J}{2}\right)\\
^{2}\Sigma_{1/2}(J\!\geq\!0)&=B_{v}(J(J+1))-D_{v}(J(J+1))^{2}-\gamma\left(\frac{J+1}{2}\right)
\end{align}
\end{subequations}

\subsection*{Triplet Levels}

The more accurate expressions for $^{3}\Sigma$ states energy
levels given by \cite{Nicolet:1989} form the formulae of
\cite{Miller:1953} with a typographical correction from
\cite{Laux:1993} are preferred to the ones found in
\cite{Herzberg:1965}:

\begin{subequations}
\begin{align}
^{3}\Sigma_{2}(J\!\geq\!2)&=B_{v}(J(J+1))-D_{v}(J(J+1))^{2}\\
&-\left(\negthickspace%
\begin{array}{l}%
\lambda_{v}-B_{v}+\frac{1}{2}\gamma_{v}
\end{array}%
\negthickspace\right)\notag\\
&-\left[\negthickspace
\begin{array}{l}%
\left(\lambda_{v}-B_{v}+\frac{1}{2}\gamma_{v}\right)^{2}
+4J(J+1)\left(B_{v}-\frac{1}{2}\gamma_{v}\right)^{2}
\end{array}%
\negthickspace\right]^{\frac{1}{2}}\notag\\
^{3}\Sigma_{1}(J\!\geq\!1)&=B_{v}(J(J+1))-D_{v}(J(J+1))^{2}\\
^{3}\Sigma_{0}(J\!\geq\!0)&=B_{v}(J(J+1))-D_{v}(J(J+1))^{2}\\
&-\left(\negthickspace%
\begin{array}{l}%
\lambda_{v}-B_{v}+\frac{1}{2}\gamma_{v}
\end{array}%
\negthickspace\right)\notag\\
&+\left[\negthickspace
\begin{array}{l}%
\left(\lambda_{v}-B_{v}+\frac{1}{2}\gamma_{v}\right)^{2}
+4J(J+1)\left(B_{v}-\frac{1}{2}\gamma_{v}\right)^{2}
\end{array}%
\negthickspace\right]^{\frac{1}{2}}\notag
\end{align}
\end{subequations}

$^{3}\Pi$ states energy levels are given by \cite{Budo:1935}:

\begin{subequations}
\begin{align}
^{3}\Pi_{2}(J\!\geq\!2)&=B_{v}
\left[%
\begin{array}{l}%
J(J+1)-\sqrt{y_{1}+4J(J+1)}\\-\frac{2}{3}\frac{y_{2}-2J(J+1)}{y_{1}+4J(J+1)}
\end{array}%
\right]\notag\\
&-D_{v}\left(J-\frac{1}{2}\right)^{4}\\
^{3}\Pi_{1}(J\!\geq\!1)&=B_{v}\left[J(J+1)+\frac{4}{3}\frac{y_{2}-2J(J+1)}{y_{1}+4J(J+1)}\right]\notag\\
&-D_{v}\left(J+\frac{1}{2}\right)^{4}\\
^{3}\Pi_{0}(J\!\geq\!0)&=B_{v}
\left[%
\begin{array}{l}%
J(J+1)+\sqrt{y_{1}+4J(J+1)}\\-\frac{2}{3}\frac{y_{2}-2J(J+1)}{y_{1}+4J(J+1)}
\end{array}%
\right]\notag\\%
&-D_{v}\left(J+\frac{3}{2}\right)^{4}
\end{align}
\end{subequations}

with

\begin{equation*}
y_{1}=Y(Y-4)+\frac{4}{3} \hspace{1cm} y_{2}=Y(Y-1)-\frac{4}{9}
\hspace{1cm} Y=\frac{A_{v}}{B_{v}}
\end{equation*}

\section{Coalesced, Weak and Strong Branches for Doublet and Triplet Transitions}
\label{sec:HLBranches}

Expressions for parallel $^{2}\Sigma\leftrightarrow^{2}\Sigma$
transitions for H\"{u}nd case $b$ are taken from
\cite{Schadee:1964}, expressions for parallel
$^{3}\Sigma\leftrightarrow^{3}\Sigma$ transitions for H\"{u}nd
case $b$ are taken from \cite{Tatum:1971}, expressions for
perpendicular $^{2}\Sigma\leftrightarrow^{2}\Pi$ and
$^{3}\Sigma\leftrightarrow^{3}\Pi$ transitions for the
intermediary case between H\"{u}nd cases $a$ and $b$ were taken
from \cite{Arnold:1969} and \cite{Budo:1937} respectively.
Expressions for parallel $^{2}\Pi\leftrightarrow^{2}\Pi$ and
$^{3}\Pi\leftrightarrow^{3}\Pi$ transitions for the intermediary
case between H\"{u}nd cases $a$ and $b$ were taken from
\cite{Kovacs:1969}.

\subsection*{Weak Rotational Branches}

For multiplet parallel transitions ($\Delta\Lambda=0$), all the
satellite rotational branches have weak strengths. For multiplet
perpendicular transitions ($\Delta\Lambda=\pm1$) the weak
rotational branches are listed in Tab. \ref{tab:NegSP}

\begin{table}[!htb]
\begin{center}
\begin{tabular}{c c c c}
\toprule
\midrule%
$^{2}\Delta'\leftrightarrow^{2}\Delta''$ & \multicolumn{3}{c}{$^{3}\Delta'\leftrightarrow^{3}\Delta''$}\\%
\midrule%
$^{O}P_{12}$ & $^{Q}P_{21}$ & $^{R}P_{31}$ & $^{O}P_{12}$\\
$^{Q}R_{12}$ & $^{R}Q_{21}$ & $^{S}Q_{31}$ & $^{P}Q_{12}$\\
$^{Q}P_{21}$ & $^{S}R_{21}$ & $^{T}R_{31}$ & $^{Q}R_{12}$\\
$^{S}R_{21}$ & $^{O}P_{23}$ & $^{Q}P_{32}$ & $^{N}P_{13}$\\
             & $^{P}Q_{23}$ & $^{R}Q_{32}$ & $^{O}Q_{13}$\\
             & $^{Q}R_{23}$ & $^{S}R_{32}$ & $^{P}R_{13}$\\
\bottomrule
\end{tabular}
\caption{\small{Weak rotational branches for doublet and triplet
transitions}}
\label{tab:NegSP}%
\end{center}
\end{table}

\subsection*{List of Coalesced Branches for $\Sigma-\Pi$
Transitions Without Spin-Splitting Effects for the $\Sigma$ State}

The coalesced rotational branches for multiplet
$\Sigma\leftrightarrow\Pi$ transitions with a neglected
spin-splitting of the $\Sigma$ state are listed in Tab.
\ref{tab:CoaSP}.

\begin{table}[!htb]
\begin{center}
\begin{tabular}{r@{}l r@{}l r@{}l r@{}l}
\toprule
\midrule%
\multicolumn{2}{c}{$^{2}\Sigma-^{2}\Pi$} & \multicolumn{2}{c}{$^{2}\Pi-^{2}\Sigma$} & \multicolumn{2}{c}{$^{3}\Sigma-^{3}\Pi$} & \multicolumn{2}{c}{$^{3}\Pi-^{3}\Sigma$}\\%
\midrule%
$^{P}Q_{12}$\,&$=P_{2}$ & $^{P}Q_{12}$\,&$=P_{1}$ & $^{T}R_{31}$\,&$=P_{1}$ & $^{R}Q_{21}$\,&$=P_{2}$\\
$^{Q}P_{21}$\,&$=Q_{1}$ & $^{Q}R_{12}$\,&$=Q_{1}$ & $^{R}Q_{21}$\,&$=P_{1}$ & $^{R}Q_{32}$\,&$=P_{3}$\\
$^{Q}R_{12}$\,&$=Q_{2}$ & $^{Q}P_{21}$\,&$=Q_{2}$ & $^{R}Q_{32}$\,&$=P_{2}$ & $^{T}R_{31}$\,&$=P_{3}$\\
$^{R}Q_{21}$\,&$=R_{1}$ & $^{R}Q_{21}$\,&$=R_{2}$ & $^{S}R_{21}$\,&$=Q_{1}$ & $^{O}P_{12}$\,&$=Q_{1}$\\
 &  &  &  & $^{O}P_{12}$\,&$=Q_{2}$ & $^{O}P_{23}$\,&$=Q_{2}$\\
 &  &  &  & $^{S}R_{32}$\,&$=Q_{2}$ & $^{S}R_{21}$\,&$=Q_{2}$\\
 &  &  &  & $^{O}P_{23}$\,&$=Q_{3}$ & $^{S}R_{32}$\,&$=Q_{3}$\\
 &  &  &  & $^{P}Q_{12}$\,&$=R_{2}$ & $^{N}P_{13}$\,&$=R_{1}$\\
 &  &  &  & $^{N}P_{13}$\,&$=R_{3}$ & $^{P}Q_{12}$\,&$=R_{1}$\\
 &  &  &  & $^{P}Q_{23}$\,&$=R_{3}$ & $^{P}Q_{23}$\,&$=R_{2}$\\
\bottomrule
\end{tabular}
\label{tab:CoaSP}%
\caption{\small{Coalesced rotational branches for multiplet
$\Sigma-\Pi$ transitions when neglecting spin-splitting for the
$\Sigma$ state}}
\end{center}
\end{table}

\end{document}